**Dynamical phenomena in the Martian atmosphere through Mars Express imaging**

A. Sánchez-Lavega[1], T. del Río-Gaztelurrutia[1], A. Spiga[2], J. Hernández-Bernal[2], E. Larsen[1], D. Tirsch[3], A. Cardesin-Moinelo[4-6], P. Machado[6]

[1]Universidad del País Vasco UPV/EHU, Escuela de Ingeniería de Bilbao, Departamento Física Aplicada, Bilbao, Spain
[2]Sorbonne Université, Centre National de la Recherche Scientifique, École Normale Supérieure, École Polytechnique Laboratoire de Météorologie Dynamique (LMD/IPSL), Paris, France
[3]German Aeroespace Center (DLR), Institute of Planetary Research, Berlin, Germany
[4]Aurora Technology for European Space Agency, ESA-ESAC, European Space Astronomy Centre, Madrid, Spain
[5]Instituto de Astrofísica de Andalucía, CSIC, Granada, Spain
[6]Institute of Astrophysics and Space Sciences, Observatório Astronómico de Lisboa, Lisbon, Portugal

**Corresponding author**: agustin.sanchez@ehu.eus
ID = 0000-0001-7234-7634

**Abstract**

This review describes the dynamic phenomena in the atmosphere of Mars that are visible in images taken in the visual range through cloud formation and dust lifting. We describe the properties of atmospheric features traced by aerosols covering a large range of spatial and temporal scales, including dynamical interpretations and modelling when available. We present the areographic distribution and the daily and seasonal cycles of those atmospheric phenomena. We rely primarily on images taken by cameras on Mars Express.

**1. Introduction**

Since the invention of the telescope, the visual observation of the planet Mars has attracted the attention of astronomers and has lead to the identification of phenomena such as albedo features, polar caps and hoods, dust storms ("yellow clouds", e.g. Gifford 1964) and white & blue clouds (James et al. 2017). Early visual observations were registered with hand-drawn images, and from the late 19th century and throughout the 20th century, photography allowed the classification of the nature of the different observed phenomena. The use of CCD-based digital cameras (since the mid-1980s) made it possible to broaden the phenomenology observed with ground-based telescopes (Martin et al. 1992). Spacecraft exploration of Mars began in 1960 and has continued to this day, using orbital vehicles, surface landers and rovers, all equipped with imaging cameras. See Snyder and Moroz (1992) for a historical review of the different missions and their achievements up to the time of the Viking missions, and James et al. (2017) to extend this review. In recent years, other orbital missions have been taking images of Mars' atmosphere in the optical range in a similar way to those described here. These are (camera/spacecraft): MARCI/MRO (since 2006, Bell et al. 2009); IUVS/MAVEN (since 2014, McClintock et al. 2015); MCC/MOM (since 2014, ended mission, Arya et al. 2015); EXI/EMM (since 2021, Jones et al. 2021); MoRIC/Tianwen 1 (since 2021, ended mission, Li et al. 2021). To these orbit-imaging capabilities, we must add the cameras on board the Hubble Space Telescope, operating since its launch in 1990.

In this paper, we present a study of dynamic phenomena in the atmosphere of Mars traced by aerosols (condensed ices and dust clouds), that are being observed in remote sensing images in the optical range from Mars Express (MEX). We will rely mainly on images obtained with the Visual Monitoring Camera (VMC) (Ormston et al. 2011; Sánchez-Lavega et al. 2018a) and the High Resolution Stereo Camera (HRSC) (Neukum et al. 2004; Jaumann et al. 2007). In short, VMC is a frame camera of 640x480 pixels covering the spectral range ∼ 400-650 nm with a Field of View (FOV) 40°x30°. The spatial resolution at nadir varies between ∼ 300 m and 12 km, allowing coverage of the entire disk and multiple images in each sequence. On the other hand, HRSC can work in push-broom mode (taking multiple images in stereo mode) or as a frame camera with a detector of 1024x1024 pixels spanning the spectral range 400-900 nm with a FOV of 0.56°x0.56° providing a spatial resolution between ∼ 2.3 m and 92 m. For some particular phenomena we will also use the information provided by the Observatoire pour la Minéralogie, l'Eau, les Glaces et l'Activité (OMEGA) (Bibring et al. 2004). VMC provides a complementary view of Mars due to its instantaneous single-field-of-view versus most common push-broom cameras, and to the highly elliptical polar orbit of MEX (∼ 300-10,000 km) versus other orbital missions with nearly circular orbits or less elliptical orbits.

## 2. Scales of dynamical phenomena

The phenomena presented here cover a wide range of spatial and temporal scales (Figure 1). The temporal range is conditioned by two main scales, the diurnal cycle (Mars' rotation period is 24.66 h) and the annual or seasonal cycle due to Mars' obliquity (25.19°) and orbital eccentricity (0.0935) (a Martian year has 668.6 sols, equivalent to 687 Earth days) (Kieffer et al. 1992). Both temporal cycles control the behaviour of aerosols in the atmosphere leading to annual cycles of dust, water and carbon dioxide ice. The shortest time scales included in this study correspond to fractions of an hour (conditioned in part by the resolution provided by series of images) whereas the longest correspond to Martian years. We note, however, that images may exhibit morphologies related to atmospheric events developing on timescales less than an hour (e.g. turbulence); yet the time resolution of imagery from orbit does not really allow following the evolution of those phenomena. Another important timescale is the radiative time constant, which represents the time taken for the temperature of an atmospheric parcel to adjust with its environment by absorption and emission of radiation. Its typical value in the Martian low and mid-atmosphere is in the range $t_{rad} \sim 0.5 - 4$ sols (Read et al. 2015; Zurek 2017).

The spatial scales covered in this work range from the smallest observable phenomena given by the maximum horizontal spatial resolution of the images $\sim 100$ m to the planetary scale $\sim \pi R_M$ with $R_M = 3390$ km the radius of Mars. In between, we classify the phenomena as pertaining to the surface layer and the Planetary Boundary Layer (PBL) (100s meters to several kilometers, Read et al. 2017), the mesoscale (10s km) and the synoptic-scale (100s-1000s km, Hunt and James 1985). This classification stems, at first, from phenomenological arguments (Orlanski 1975) but theoretical considerations on relevant spatial scales for atmospheric events offers a justification for the definition of these typical spatial scales. For instance, an important scale for synoptic-scale dynamical processes is the Rossby deformation radius defined as $L_D = NH / f \sim 720$ km, where $N \sim 0.67 \times 10^{-2}$ s$^{-1}$ is the Brunt-Väisälä frequency, $H \sim 10.8$ km is the scale-height and $f \sim 1.0 \times 10^{-4}$ s$^{-1}$ the Coriolis parameter (Zurek 2017). A detailed discussion and evaluation of the range of values of the spatial scales (Figure 1) in relation to $L_D$, especially in comparison with the Earth, can be found in Rafkin et al. (2017). The scale-height $H = R * T / g$, where $R* = 192$ J kg$^{-1}$K$^{-1}$ is the gas constant for Mars atmosphere and $g = 3.72$ ms$^{-2}$ the acceleration of gravity, is also a relevant scale in vertical motions.

Complementarily, a widely used dynamic classification of weather systems is based on the Rossby ($Ro$) and Richardson ($Ri$) numbers, defined as $Ro = u / fL$ and $Ri = N^2 / (\partial u / \partial z)^2$ respectively (Emanuel 1986). Here $u$ is a representative horizontal velocity and $L$ the corresponding spatial scale; $N$ is again the Brunt-Väisälä buoyancy frequency at which a displaced air parcel will oscillate when displaced vertically within a statically stable environment. Broadly speaking, because Mars and Earth are rapidly rotating planets, these numbers depend mainly on latitude ($\varphi$) and altitude ($z$), since to first order $u$ ($\varphi, z$) and $N(z)$, and $f = 2\pi \sin \varphi$. Values of $Ri < 0$ ($Ri \sim -1$, $N^2 < 0$) are related to the development of convective instability, when the static stability $S= N^2g/T < 0$ (surface layer and PBL). This occurs for daytime convective turbulence in the Planetary Boundary Layer and, at higher altitudes, for gravity wave breaking. The range $0 < Ri < 1/4$ corresponds to small-scale turbulent processes dominated by vertical wind shear. These low values of the Richardson number take place on scales $< H$ near the surface and in the PBL where friction forces play an important role, or in thin layers in the free atmosphere. Values of $Ri \sim 1$ imply a balance between static stability and vertical wind shear favoring the development of the inertial instability. Then the motions tend to be aligned with the wind shear vector. At higher altitudes ($z > H$), $Ri >> 1$ due to the large values of $N$, and baroclinic instability develops at mid and high latitudes ($\varphi \sim 40°$ to $80°$ on Mars) where $Ri \sim 5 - 100$. Large-scale motions with $L \sim R_p$ (planet's radius) such as Hadley cell circulations, occur for even larger values of $Ri$. An excellent discussion of the dominant regimes for this range of values of the Richardson number can be found in Stone (1976) and, for Mars, in Rafkin et al. (2017).

The Rossby number compares the advection time scale $L/u$ by a wind of velocity $u$ and distance $L$ to the planetary rotation temporal scale $1/f$ (i.e. flow to Coriolis accelerations). Low values of Ro < 1 imply geostrophic balance between the Coriolis force and the pressure gradient as dominant terms in the momentum equation, a situation we find in the mid and subpolar latitudes of Mars. When $Ro \sim 1$ the centrifugal force intervenes in the motion and the regime is in gradient balance. Very high values of $Ro \gg 1$ imply that the Coriolis force plays no role in the motions, as in equatorial region where $f \sim 0$. Planetary and synoptic scale motions in geostrophic and quasi-geostrophic balance tend to be two-dimensional ($L \gg H$) whereas in the mesoscale regime the motions are three-dimensional ($L \sim H$) and neither the flow acceleration nor the Coriolis force can be neglected ($Ro \sim 1$) (Rafkin et al. 2017).

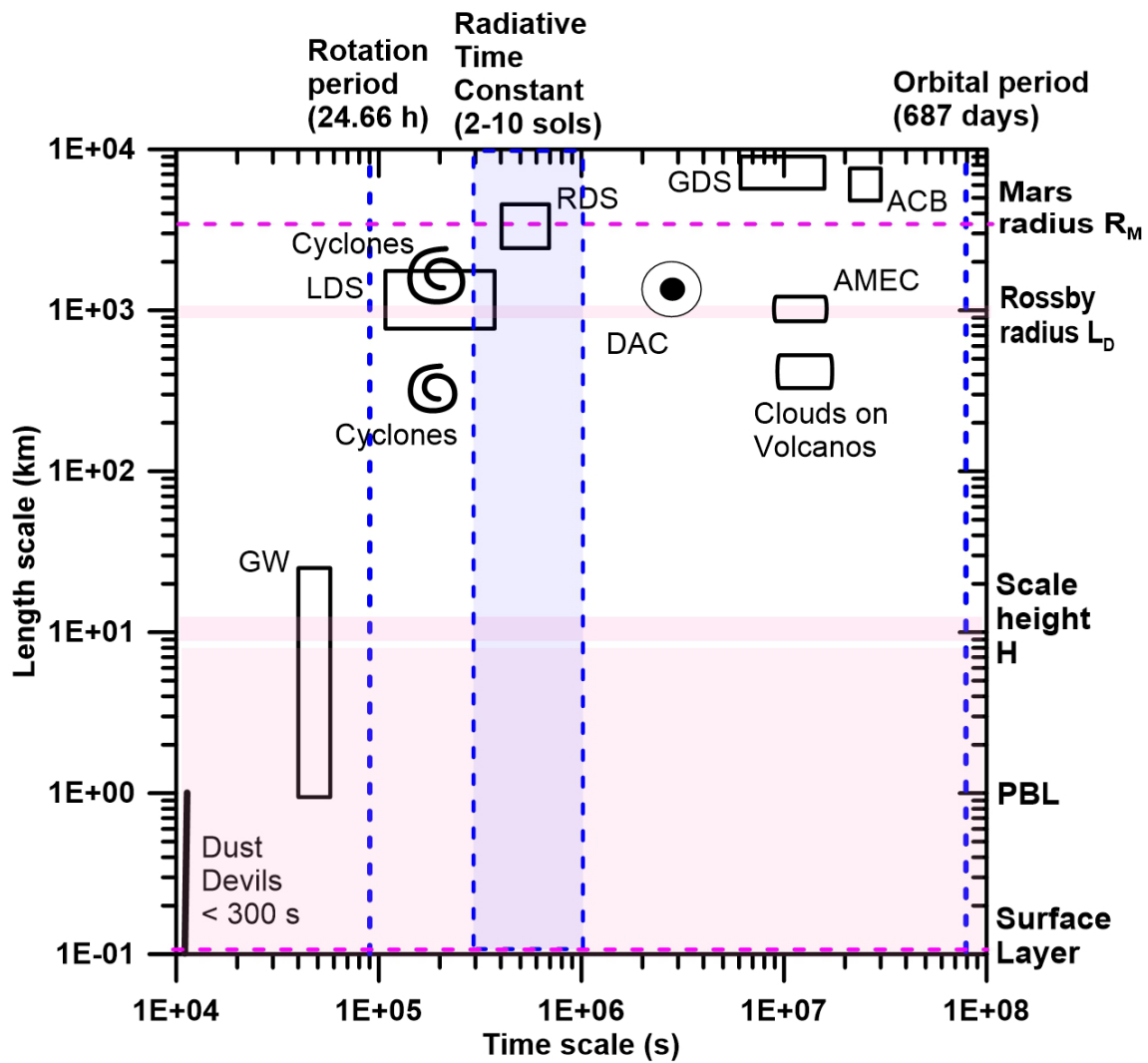

**Figure 1.** *Mars dynamical features identified in a temporal - spatial scale plot from images taken from orbit. The acronyms are GW (gravity waves), ACB (Aphelion Cloud Belt), DAC (double-annular cyclone), AMEC (Arsia Mons Elongated Cloud). The Dust Storms acronyms correspond to Local (LDS), Regional (RDS) and Global (GDS). Dust Devils are below the temporal scale (representative lifetimes < 5-30 min). Three main temporal scales are shown with blue dotted vertical lines: the Martian day or sol, the radiative time constant and the Martian year. All relevant spatial (length) scales are represented*

*by magenta colour: the surface layer (dotted line), Planetary Boundary Layer – PBL (broad horizontal band), the scale height H (narrow horizontal band), the Rossby deformation radius $L_D$ (horizontal band) and the Mars radius $R_M$ (dotted line).*

## 3. Seasonal and areographical distribution of the phenomena

The three time scales indicated in the previous section (solar day, radiative time constant and orbital period or year) govern the cyclic behaviour of the aerosols in the atmosphere. Figure 2 shows the seasonal dependence of the meridional distribution of atmospheric phenomena traced by aerosols. As it is well known, the first half of the year ($L_s \sim 0°$-$180°$) corresponds to the period of prominent water cloud ice development, forming the Aphelion Cloud Belt (ACB) that extends along all longitudes in the tropical area. Dust storms also occur mainly at the edges of the polar caps. The second half of the year ($L_s \sim 180°$-$360°$) corresponds to the dusty season, when dust storms of different sizes form frequently, with specific areas of occurrence distributed throughout the geography of the planet (Kass et al. 2016; Barnes et al. 2017; Battalio and Wang 2021).

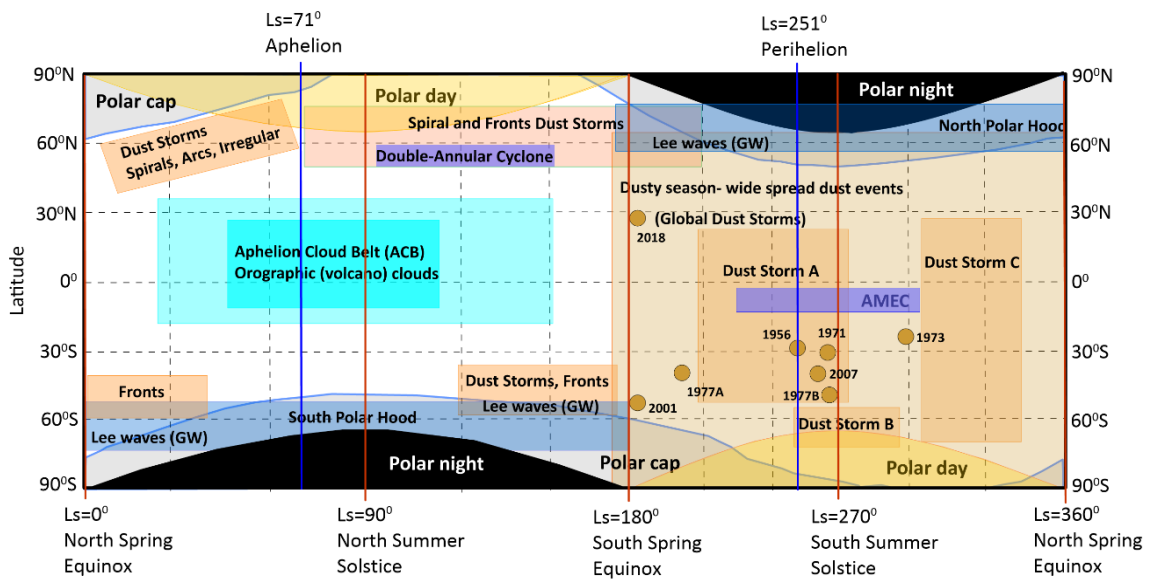

**Figure 2.** *Main dynamical phenomena on Mars seen by condensate clouds and dust in summarizing all spacecraft images taken from orbit as represented on a Latitude – Solar Longitude ($L_s$) plot. The color rectangular areas mark the approximate range of maximum development and meridional extent of the phenomena indicated. The brown dots indicate the onset of the Global Dust Storms (GDS) identified by the terrestrial year of their development. The beginning of each season, aphelion and perihelion dates are indicated by red and blue lines, respectively. The polar day and night periods are also shown.*

Figure 3 presents the aerographic distribution of these phenomena. The geological dichotomy of Mars conditions the development of some of the atmospheric phenomena. The northern hemisphere is characterised by the great Planitias (Arcadia,

Acydalia and Utopia Planitia) where dust storms develop, together with the volcanic areas of Tharsis, Elysium and Alba Patera, home of orography forced dynamics. The southern hemisphere, location of dust storm development, is dominated by the abrupt territory of numerous impact craters, including the two large basins of Argyre and Hellas. Finally, the edges of the polar ice caps are the sites for the development of dynamic instabilities in the dominant zonal flow produced by strong meridional temperature gradients. In particular, the northern hemisphere is the site of abundant spiral-shaped cyclonic vortices.

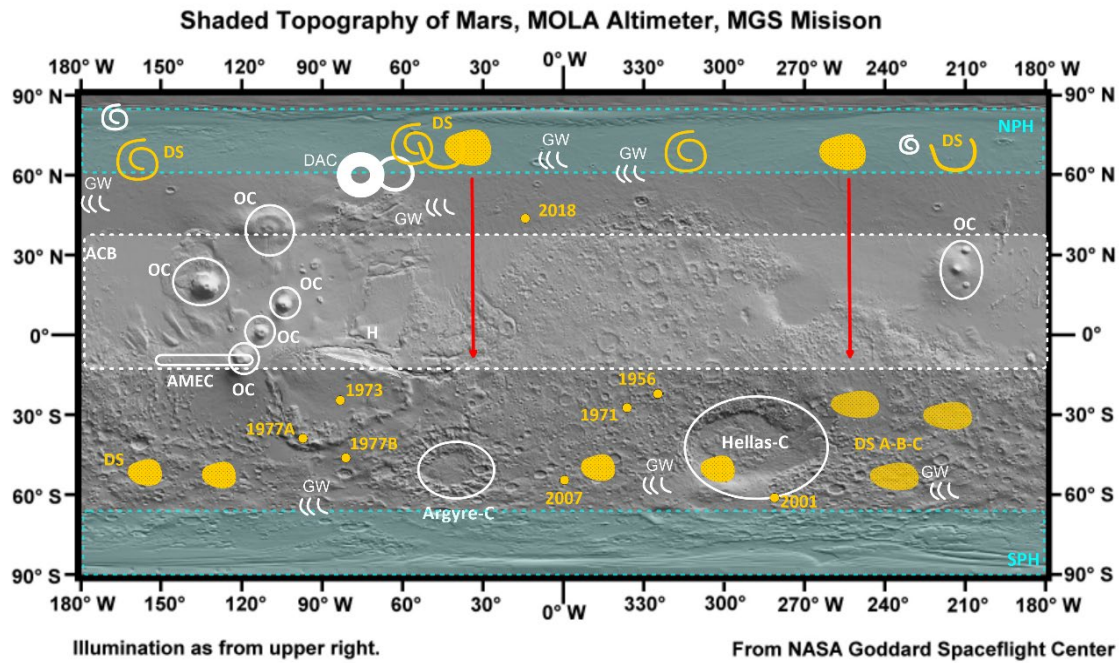

**Figure 3**. *Main dynamical phenomena on Mars identified on a Latitude – Longitude topographic map. The features are indicated by their acronym and different symbols. DS (Dust Storms) are indicated in yellow color and their different characteristics are represented by schematic spirals, arcs (flushing storms, Wang et al. 2023) or a dotted area (irregular and textured storms). The location where A, B, C storms mainly evolve in the southern hemisphere is indicated. The location of the onset of Global Dust Storms is identified by dots, with the terrestrial year of their development next to them. In white, spirals denote cyclones traced mainly by clouds, and the recurrent annular-double cyclone (DAC) is represented by a double white ring (the main cyclone indicated with a thicker line). Orographic clouds (OC) over volcano areas are marked by white circles, the AMEC by the elongated rectangle and lee waves (gravity, GW) by three small white arcs. The letter H represents Hazes on Valles Marineris and circles surrounding Argyre and Hellas craters, hazes in those regions. The clouds forming the North and South Polar Hood (NPH, SPH) are marked by the light blue areas within dotted blue lines. The Aphelion Cloud Belt (ACB) is indicated by the light white area within dotted white lines. The red arrow indicates the corridor through which dust storms migrate.*

## 4. Planetary-scale phenomena

In this section, we focus on three specific large-scale phenomena traced by clouds and dust: The tropical aphelion cloud belt (ACB), the major Global Dust Storms (GDS) or Planetary Encircling Dust Storms (PEDS), and the dynamics in the polar regions (the North and South Polar Hoods and polar vortices).

## 4a. The Aphelion Cloud Belt (ACB)

The ACB is an irregular cloud band formed by water ice clouds that develops in the tropical area of Mars between latitudes $\sim$ 10°N and 30°S during the northern hemisphere spring and summer seasons. It reaches its maximum zonal extent between $L_s \sim 80°$ and 150° (Clancy et al. 1996; Wang and Ingersoll 2002; Clancy et al. 2017; Wang and González Abad 2021). During the ACB season, clouds also form over high altitude volcanic regions (Olympus, Ascraeus, Pavonis, Arsia), further north of the tropics (Elysium and Alba Patera areas) and over depressed areas further south (Hellas Planitia) (Figure 4). Other regions of frequent cloud cover during that season are Valles Marineris and Syrtis Major Planum.

The formation of the ACB is due to the low temperatures (aphelion is at $L_s = 71°$) and water vapor abundance during this season, resulting from water vapor transport by large-scale circulations, which produce clouds with optical depths $\sim$ 0.2-06 (in UV and violet wavelengths) that reach top heights $\sim$ 20 km. As Mars approaches the North autumn equinox ($L_s = 180°$), the increasing temperatures and dust lifting warm the atmosphere and cloud formation declines. The ACB shows large diurnal morning-evening variability and zonal and meridional diurnal trends (Wolff et al. 2019). Other spatial and temporal variability observed in the ACB can be due to the presence of an eastward propagating equatorial Kelvin wave (zonal wavenumber 1 and period between $\sim$ 20-30 sols) (Wang and González Abad 2021).

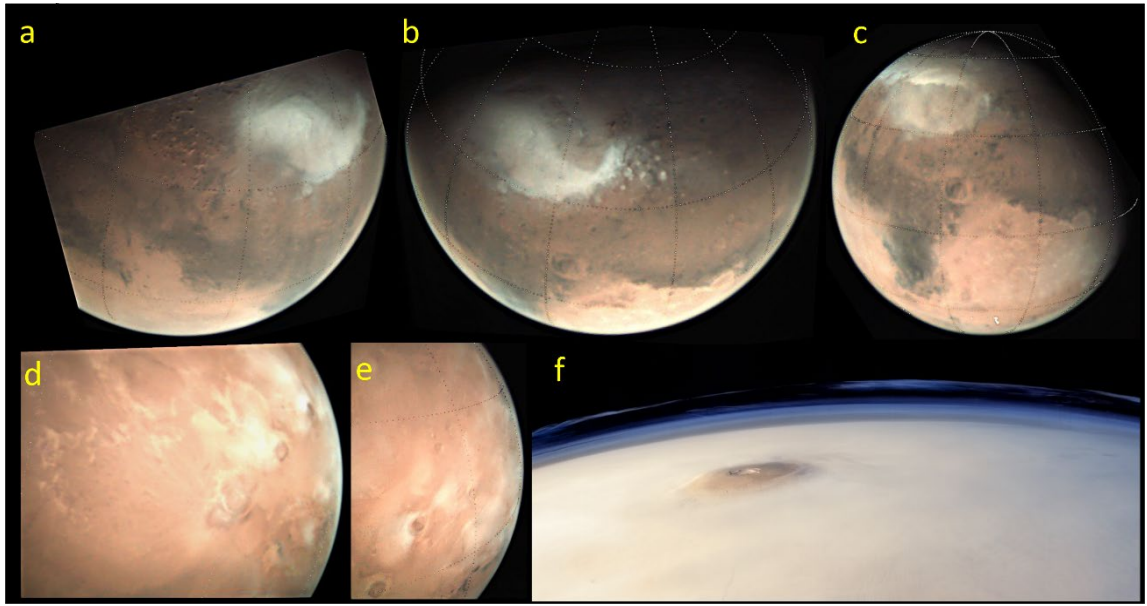

**Figure 4.** *Images of water ice clouds forming on orographic features during the development of the ACB. Changing clouds on Hellas basin (probably mixed with surface frost) are seen on (a) 1 September 2021 ($L_s$ = 93°); (b) 22 September 2021 ($L_s$ = 102°); (c) 17 December 2021 ($L_s$ = 143°). South is up and North is at the bottom. Clouds over the Tharsis volcanos are seen in (d) 25 September 2021 ($L_s$ = 103°); (e) 13 November 2021 ($L_s$ = 102°); (f) Ascraeus Mons on 16 April 2016 ($L_s$ = 131°). Images (a)-(e) from VMC and (f) from HRSC, processed by Aster Cowart.*
(*https://www.flickr.com/photos/132160802@N06/30564409838*)
*Credits: ESA/DLR/FU Berlin/J. Cowart, CC BY-SA 3.0 IGO*

## 4b. The Global Dust Storms

Global Dust Storms (GDS) or Planet Encircling Dust Storms (PEDS) are exceptional events during which the entire surface of Mars is covered with dust for a period of 2-3 months (Kahre et al. 2017; Shirley et al., 2020). The column integrated dust opacity at 880 nm reached during the 2018/MY 34 GDS a peak value ∼ 8.5 (Guzewich et al. 2019). They are rare phenomena and difficult to predict, since only eight confirmed cases have been observed in the last 62 years (1956, 1971, 1973, 1977A, 1977B, 2001, 2007 and 2018). They occur during the perihelion season (southern spring and summer), when, due to the eccentricity of the orbit, an annual maximum of insolation occurs in the southern hemisphere.  The earliest in the Martian year were those in 2001 and 2018 ($L_s$ ∼ 180°) and the latest the one in 1973 ($L_s$ ∼ 300°) (Sánchez-Lavega et al. 2019). The 2018 GDS is the only one that is confirmed to initiate in the northern hemisphere (Sánchez-Lavega et al. 2019). For the other cases, the onset occurred in the southern hemisphere between latitudes 30°S and 60°S and in a 160°-wide band in longitude, between 90°W and 285°W (Montabone et al. 2020), although Wang and Richardson (2015) mentioned that the 2005/MY 28 GDS might have started from the northern hemisphere. The onset of the 2018 event took place at mid-latitudes of the northern hemisphere in Acidalia Planitia, expanding rapidly with peak speeds in the range of 14 to 40 ms⁻¹ in the eastward and southward directions (Sánchez-Lavega et al. 2019). The explosive onset and rapid expansion of a GDS versus the usual behavior of local or regional dust storms has been proposed to be the result of a radiative-dynamical coupling produced by increased dust heating within the Hadley circulation, leading to enhanced flow convergence and intensified motion and larger spread of dust (Kahre et al. 2017).

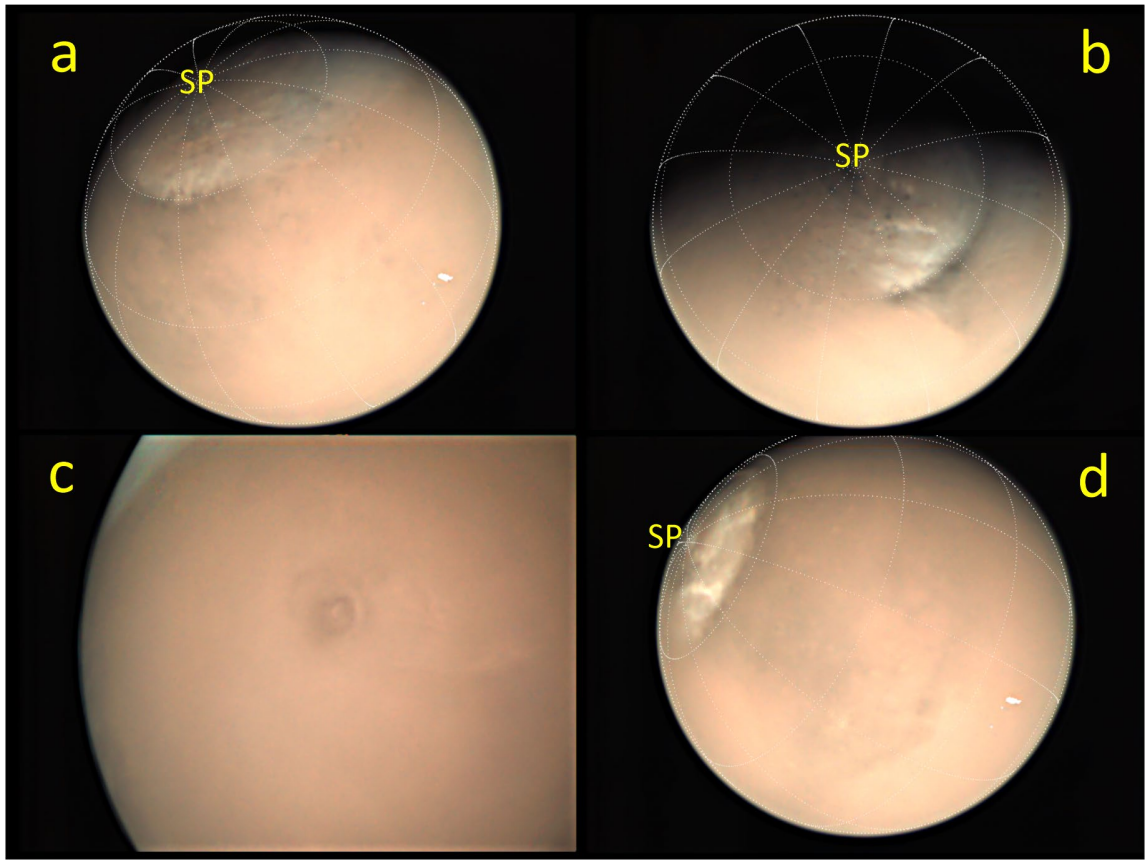

**Figure 5.** *Images of GDS 2018 by VMC. (a) Southern hemisphere and South Pole (SP) on June 23 (centre 142°E, 41°S) $L_s$ = 198.4°; (b) South Pole partially covered with dust and surroundings on June 24, $L_s$ = 199.1°; (c) Northern hemisphere with Olympus Mons emerging from the dust on 2 July, $L_s$ = 203.8°; (d) Southern hemisphere and South Pole on July 23 (centre 8°E, 25°S) $L_s$ = 216°.*

### 4c. Polar dynamics

The dynamics of the polar regions is strongly conditioned by the seasonal cycle of evaporation and condensation of carbon dioxide at both poles, and consequently by the huge change in the extent of the polar ice caps (Figures 1 and 2). Due to the low temperatures prevalent during the polar night, an immense cloud cover develops over the poles throughout their respective autumn and winter, known traditionally as North and South Polar Hoods (NPH and SPH) (Clancy et al. 2017). The cloud cover of the NPH is more extended, rich in morphology and heterogeneity in optical depth, extending at its maximum down to latitudes ∼ 45°. The structure and evolution of the polar hoods depend more on the atmospheric temperature variations than in water-vapor abundance, or its variability in altitude produced by thermal tides (Benson et al. 2010). The circumpolar hazes in the polar hoods are cirrus-like water-ice clouds, extending from the surface up to 50 km. During the polar winter and poleward of ∼ 60° the lower clouds (0-15 km) at the polar cap are formed by $CO_2$ ice particles (Clancy et al. 2017).

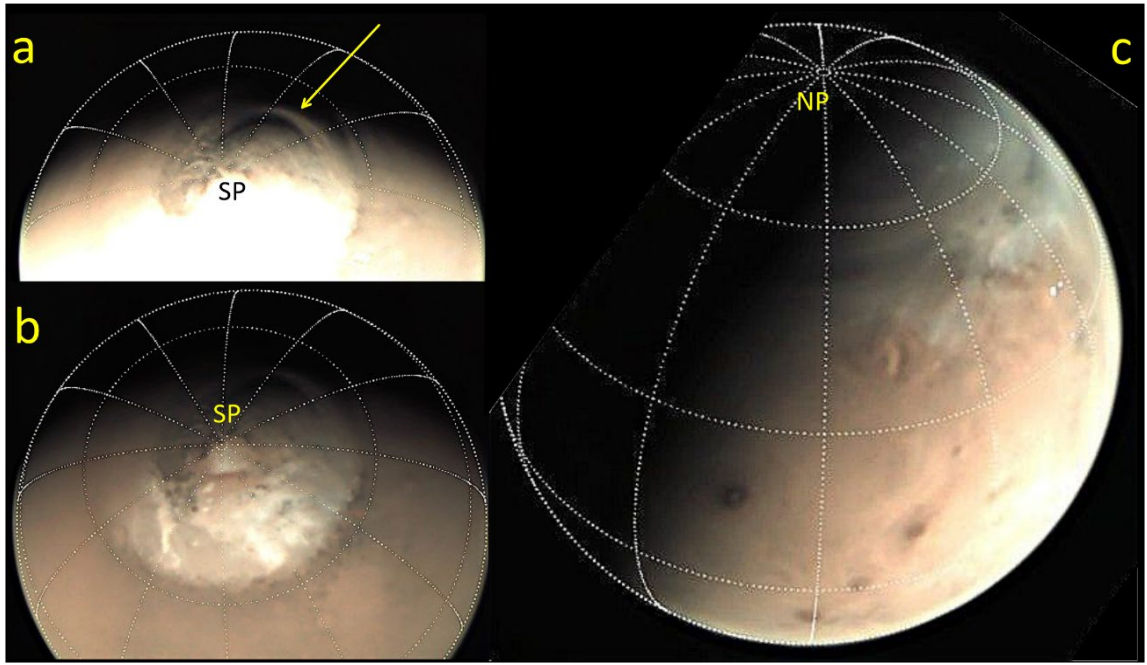

**Figure 6.** *Images from VMC/MEX showing the Polar areas with dust and clouds. South Pole into view: (a) and (b) Images obtained on 22 July 2018 during the development of the Global Dust Storm GDS 2018 (Ls = 216°) with different exposure times to show an arc of dust penetrating (marked with an arrow) and surrounding the South pole (a). North Pole into view: (c) Image obtained on 1 November 2022 (Ls = 331°) showing the NPH clouds reaching latitude 45°N.*

The polar atmospheric dynamics is dominated by the development of a polar vortex which is traced by an annulus of high potential vorticity (PV) (Banfield et al. 2004). It has been proposed that the latent heat associated with $CO_2$ condensation in the polar winter regions is a driver for the formation of this annular PV structure (Toigo et al. 2017; Ball et al., 2021). In the absence of a stratosphere, the Martian polar vortex is strongly controlled by the global-scale Hadley circulation, forming on the poleward side of the descending branch. Changes in the Hadley circulation driven by atmospheric dust variability, can therefore significantly affect the strength and structure of the polar vortex (Mitchell et al. 2015; Seviour et al. 2017). For example, dust penetrated massively in the South polar region following the development of the Global Dust Storm in 2018 (Sánchez-Lavega et al. 2019). The dust was organized in long filamentary bands (length ∼ 3000 km) that surrounded the South pole, strongly modifying its circulation (Hernández-Bernal et al., 2019) (Figure 6).

## 5. Synoptic-scale phenomena

In this section, we consider as synoptic-scale those weather systems occurring at spatial scales in the range between ~ 100 − 2,000 km and with typical duration of a few sols (Figure 3) (Chamberlain et al. 1976; Hunt and James 1985).

## 5a. Cyclone systems

Spiral-shaped systems traced by dust and sometimes accompanied by water-ice clouds, are abundant in the northern hemisphere of Mars but scarce or absent in the southern hemisphere (Wang and Ingersoll 2002). These features are very abundant at the edge of the north polar cap (NPC) and their distribution in latitude follows the recession of the polar cap during northern spring ($L_s \sim 0°$-90°). They are less frequent in summer time ($L_s \sim 90°$-180°) (Figure 7). They locate approximately between latitudes 55°N and 75°N with an important concentration in in Acidalia Planitia, in the longitude sector from 270° to 360°E and 0°-30°E (Clancy et al. 2017; Kahre et al. 2017). They change rapidly each sol, diluting at dusk and forming again at dawn, reaching their maximum development at about noon, and their centroids move with velocities in the range 5-30 ms$^{-1}$ in the northeast direction. The dust spirals have sizes in the range L ~ 500-2000 km and sometimes show in their interior well-defined areas of textured patterns (grainy and blob-shaped), typically 5-15 km in size, suggestive of dry-convective motions (Sánchez-Lavega et al. 2022a).

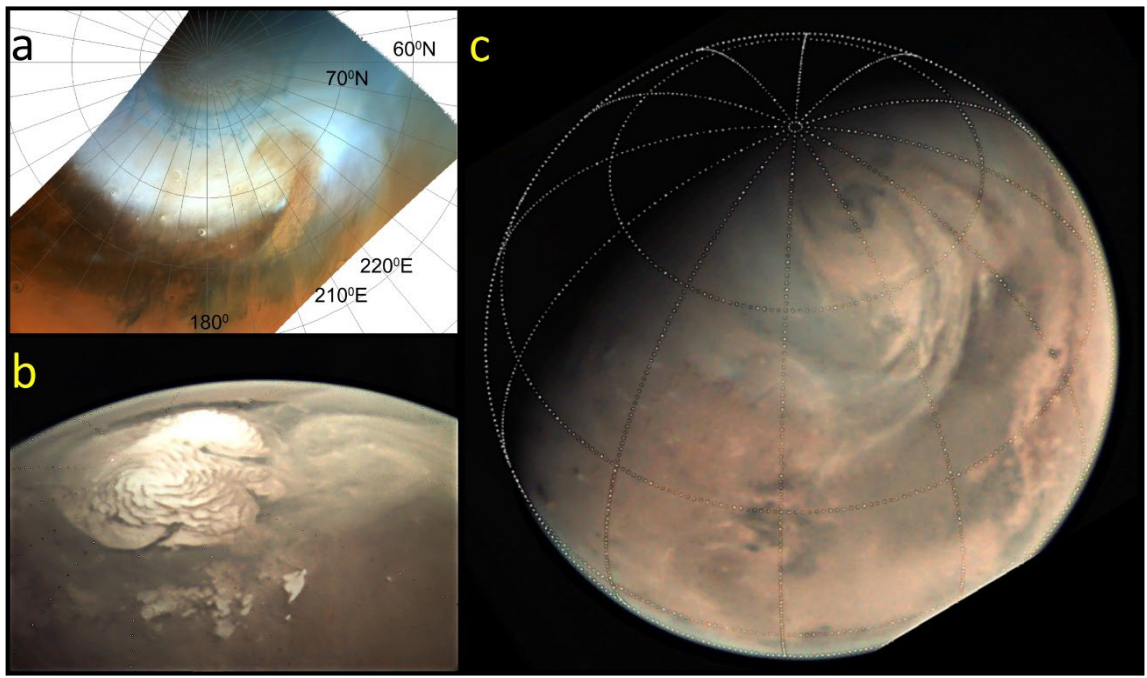

**Figure 7.** *Spiral cyclones around the edge of the North Polar Cap. (a) A dust cyclone mixed with water-ice clouds imaged by HRSC on 26 May 2019 ($L_s$= 30°). See Sánchez-Lavega et al. (2022a); (b) A dust spiral imaged by VMC on 4 October 2019 ($L_s$= 88°); (c) A large dust spiral imaged by VMC in 1 October 2012 ($L_s$= 180.7°).*

As cloud generation progresses in the Aphelion Cloud Belt (ACB), spiral systems of condensate clouds form in the same range of northern latitudes, first observed by Viking orbiters ($L_s \sim 105°$-$125°$) (Gierasch et al. 1979; Hunt et al. 1979). The most iconic representative is the double-annular cyclone (DAC), so called because of the shape traced by the water ice clouds that form it (Cantor et al. 2002; Sánchez-Lavega et al. 2018b) (Figure 8). Similarly to the spirals described above, this seasonally recurrent cyclone forms in the morning and dissipates at night but regenerates each day during the season of occurrence ($L_s \sim 115°$-$135°$). It evolves in the Acidalia Planitia (longitude sector 245°E - 30°E, centre latitude 60°N) translating eastward with velocities $\sim$ 1-4 ms$^{-1}$. The clouds organize in a nearly circular pattern of radius 600-800 km around a central cloud-free core of radius $\sim$ 100-300 km, revealing the prevailing cyclonic circulation of the system, with typically tangential velocities $\sim$ 5-20 ms$^{-1}$.

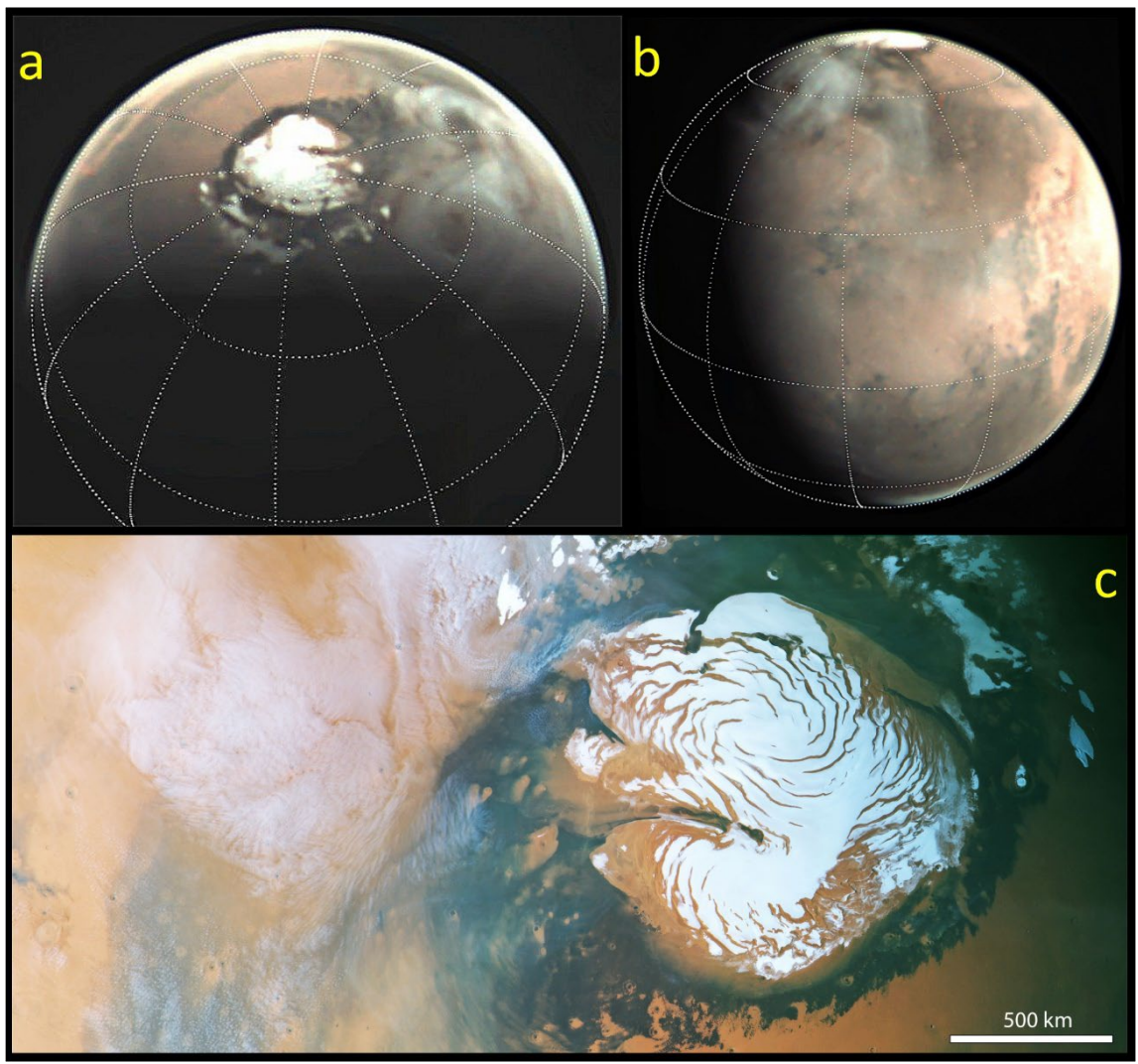

**Figure 8**. *The annually recurrent double-annular cyclone (DAC) formed by water ice clouds at the edge of the North Polar Cap. (a) VMC image on 18 June 2012 (MY 31, $L_s$= 126°) showing the two vortices forming in the morning; (b) VMC image on 21 March 2016 (MY 33, $L_s$= 125.6°) showing the leading vortex; (c) HRSC image on 2 May 2014 (MY 32, $L_s$= 130.4°) showing fine details of the cloud structure.*

The extratropical and nearly polar latitude band where these cyclones occur is characterized by a large meridional temperature gradient that drives a vertically-sheared westerly jet stream, where intense transient eastward traveling waves grow as baroclinic and barotropic disturbances (Sánchez-Lavega et al. 2018b for DAC; Sánchez-Lavega et al. 2022b for dust spirals). Tyler and Barnes (2014) highlighted through mesoscale modeling how this synoptic-scale environment interacts with topographically-induced circulations and western boundary current effects to give rise to the transient eddies observed in satellite imagery on Mars.

**5b. Local and Regional dust storms**

In addition to the global dust storms (GDS) described in section 4b and the cyclonic systems in section 5a, there are other storms with a variety of sizes, shapes and textures (Battalio and Wang 2021). According to their area $A$, storms are classified as local (A < $1.6X10^6$ km$^2$) and regional (A > $1.6X10^6$ km$^2$) (Cantor et al. 2001). They exhibit two basic configurations: (a) frontal, e.g. "flushing" or arc-shaped and "comma" storms. (b) irregular system, most of them with well-defined textured patterns (Cantor et al. 2001; Wang et al. 2003, 2005, 2007; Wang and Fisher 2009; Wang et al. 2011; Hinson and Wang 2010; Wang and Richardson 2015; Guzewich et al. 2015; Kahre et al. 2017; Kulowski et al. 2017; Sánchez-Lavega et al. 2022). Figure 9 shows some examples of this kind of dust storms as imaged by VMC during the non-dusty season (Sánchez-Lavega et al. 2018a, 2022).

During the dusty season (see Figure 1) both hemispheres show similar rates of dust storm formation (Kahre et al. 2017, see their fig. 10.9). In the north, they are abundant in the Arcadia, Acidalia and Utopia Planitia particularly at $L_s \sim 180°$-$240°$ and from $305°$ to $30°$ (Figures 2-3). In the Southern Hemisphere, the prevailing type of local dust storms have irregular shapes and, as in the north, most of them exhibit textured patterns (Battalio and Wang 2021; Battalio et al 2023; Wang et al. 2023). They grow at the edge of the South Polar Cap (SPC) in the 50°S-70°S latitude band, with the Hellas Planitia being one of the most productive regions for storm formation. During years without a GDS, there are three regional-scale dust storms, known as A, B and C, that originate in the southern hemisphere (Kass et al. 2016). These storms tend to repeat seasonally starting sequentially at $L_s \sim 225°$ (storm A), $\sim 250°$ (storm B) and $\sim 310°$ (storm C), expanding along the northern latitudes but propagating (flushing) in some occasions into the southern hemisphere where they subsequently grow (Wang and Richardson 2015; Battalio and Wang 2021).

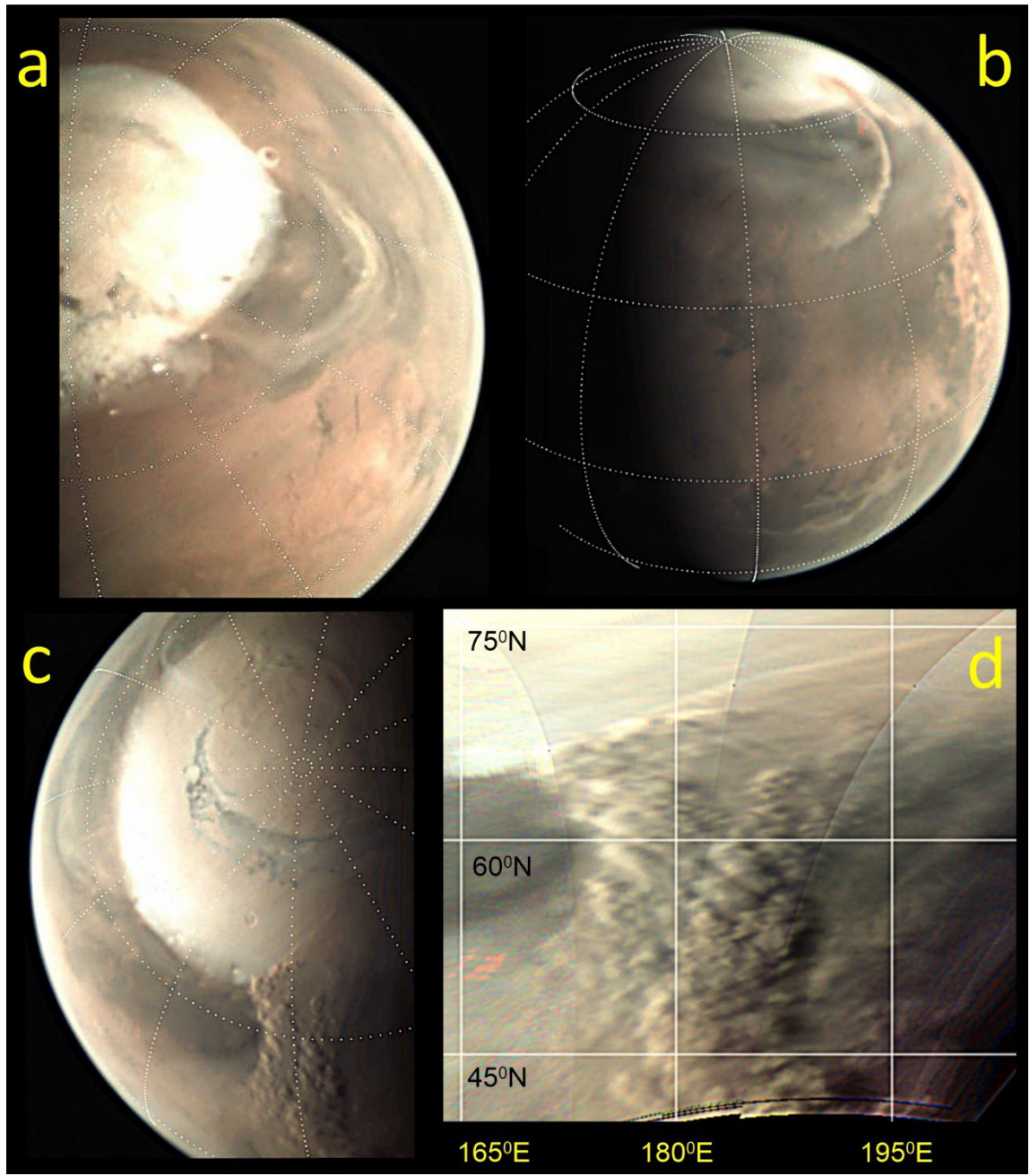

**Figure 9.** *Arc-shaped and textured local dust storms observed with VMC at the North Pole edge. (a) Arc storm in the NPC edge on 9 July 2019 (312°E, 63°N, 11.2 hr LTST, $L_s$ = 50.2°, MY 35); (b) Arc-frontal storm on 30 March 2021 (310°E, 55°N, $L_s$ = 24.8°, 8.4 hr LTST, MY 36); (c)-(d) Textured dust storm on 29 May 2019 (185°E, 55°N, 16-7 hr LTST, $L_s$ = 32°, MY 35).*

Although dust storms are less frequent in the equatorial region than in the polar cap edges (Kahre et al. 2017), they are of high interest because most surface space missions (rovers and platforms) operate at or near the equator (Figure 10). An example of one such storm that evolved at the location of the Perseverance and Curiosity rovers is shown in Figure 10b. Another remarkable storm event is the one witnessed at tropical latitudes, with particularly high optical depths, by Määttänen et al. (2009) using OMEGA spectro-imagery; this storm is thought to host convective motions triggered by the absorption of incoming sunlight by dust particles within the storm (see simulations by Spiga et al. 2013).

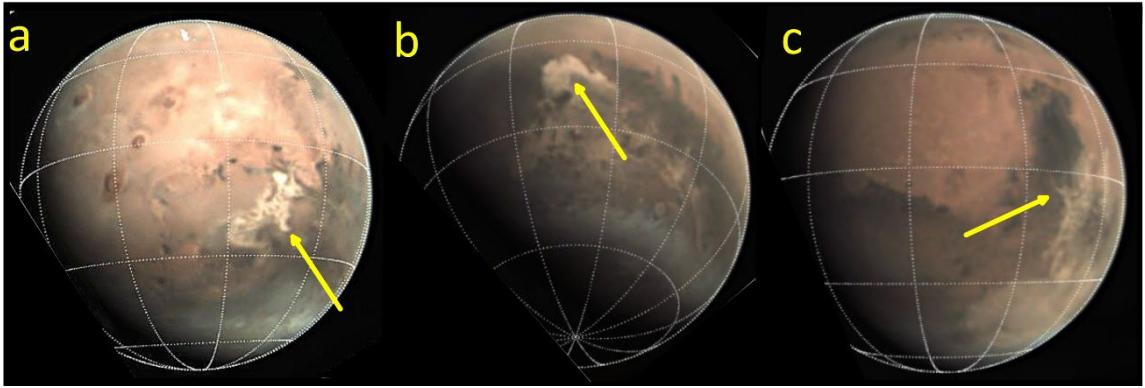

**Figure 10.** *Equatorial irregular dust storms by VMC in MY 36 marked by arrows. (a) 21 December 2021 (storm center at 282.4°E, 18.05°S, 10.7 LTST, L$_s$ = 145°); (b) 01 January 2022 (storm center at 105.7°E, 14.6°S, 8.9 LTST, L$_s$ = 151°); (c) 04 January 2022 (storm centre at 77.5°E, 8°S, 11.3 LTST, L$_s$ = 152°). North is up and East to the right in these images.*

## 6. Mesoscale phenomena

We consider in this section those phenomena with spatial scales in the range of ~ 1 - 1000 km, most typically ~ 10 - 100 km, that is, scales approximately ranging from the vertical scale height (H) up to the mid-synoptic scale (see Figure 1, Emanuel 1986, Rafkin et al. 2017). We also include dust devils, which are under the control of the surface layer and the PBL and have sizes below that range since dust devils are caused by turbulent vortices (Read et al. 2017).

### 6a. Orographic features

The interaction of the atmospheric flow and Martian orography (volcanoes, craters, basins) produces different kinds of disturbances that can be traced both by the dust distribution and by the presence of water-ice clouds of orographic origin. Orographic cloud formation is very active during the ACB phase (Figure 4 and section 4a) and can be triggered by two dynamic mechanisms: (a) Water enrichment due to the vertical transport of water vapor by winds over topographic slopes, (b) Temperature drops due to vertical propagation of gravity waves. Both the enrichment of water by thermal winds and the temperature drop by gravity waves can act together to achieve cloud condensation conditions, which depend on the water content and temperature of the unperturbed environment (Michaels et al. 2006; Hernández-Bernal et al. 2022).

### *Clouds formed by volcanos*

The clearest cases of orographic clouds on Mars are present in the volcanic area of Tharsis, where the four biggest volcanoes on Mars are located. During the aphelion season, near condensation conditions are present at low altitudes in the tropical latitudes forming the ACB, and anabatic winds transport water and aerosols through the slopes of volcanoes (Michaels et al. 2006) (Figure 4). The enrichment of water due to vertical transport is enough to drive the formation of clouds, and the predominant westward winds in this season (at the latitudes of Tharsis) blow lifted water and dust to form long the cloud plumes usually present in images of Tharsis (Michaels et al.. 2006).

During the perihelion season, global average temperatures are higher, and near condensation conditions in tropical latitudes are only present at higher altitudes (Fedorova et al. 2021). Consequently, forming orographic clouds requires that perturbations reach higher altitudes. In 2021, Hernández-Bernal et al. (2021a) reported a striking case of an orographic cloud expanding from Arsia Mons. This Arsia Mons Elongated Cloud (AMEC) develops during the perihelion season in the range of $L_s \sim 220°$-320° (Figures 2-3), and exhibits an unusually long tail, which grows during a few hours after sunrise and quickly evaporates before the afternoon, in a regular daily cycle that repeats from sol to sol (Figure 11). Measurements in MY 34 revealed that the cloud tail grew at a velocity of 170 ms$^{-1}$ reaching lengths of up to 1800 km. Mesoscale modeling, run at $L_s = 270°$, MY 34, showed the presence of a cold pocket as described above, 30K below the environment, which explains the daily evolution of the head of the cloud over the western slopes of Arsia Mons (Hernández-Bernal et al. 2022a). Reproducing the long tail of the AMEC is still a challenge for models.

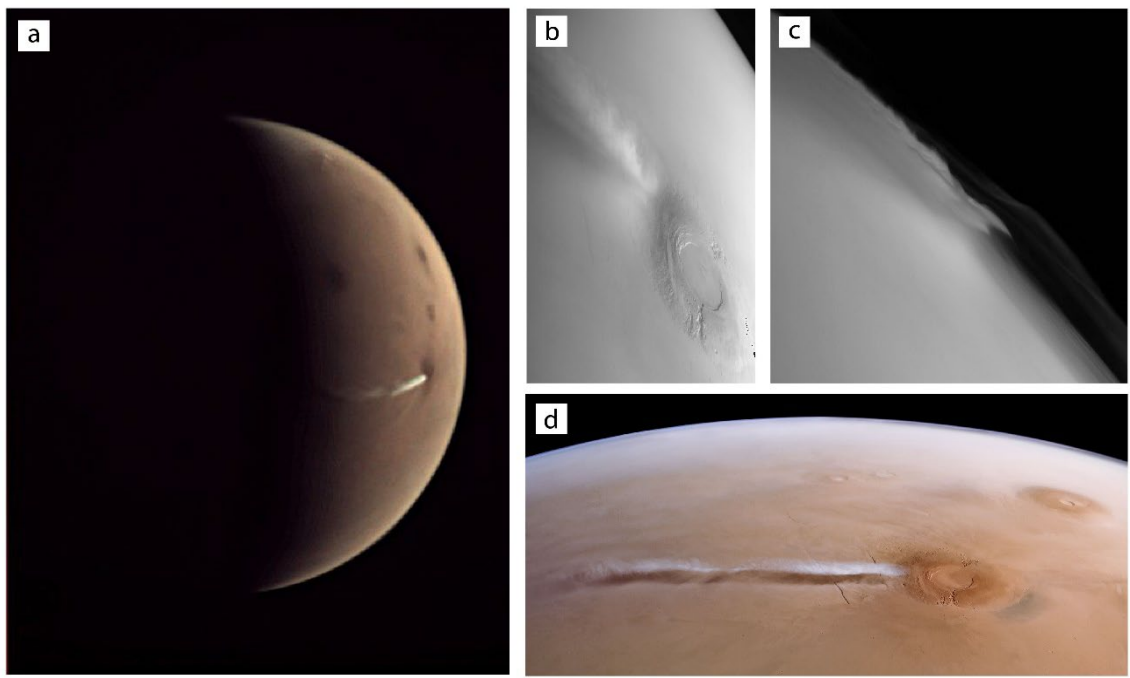

**Figure 11.** *Four views of the AMEC phenomenon by VMC and HRSC. (a) Image taken by VMC on 19 July 2020 (8 hr LTST, $L_s = 241°$) showing the AMEC extending to the West from Arsia Mons (North is up and East is at right); (b) Nadir channel image of HRSC limb observation 20922_0000 taken on June 20th, 2020 from an altitude of about 2800 km with the AMEC likewise extending to the West. (ESA/DLR/FU Berlin) (c) Red channel image of the same HRSC limb observation as panel (b) that "looks" inertially to the limb showing the vertical structure of the cloud. (ESA/DLR/FU Berlin) (d) RGB-color composite*

*of the high-altitude HRSC observation 18627_0000 showing the AMEC on September 9th, 2018. Credits: ESA/DLR/FU Berlin/J. Cowart)*

### *Elongated clouds*

A very common phenomenon of orographic origin on Mars is the presence of narrow and elongated water-ice clouds, with the AMEC described above perhaps the most paradigmatic case. Other elongated clouds, however, may not be related to an orographic origin (or the source is unclear).

Cloud trails were reported in Valles Marineris by Clancy et al. (2009, 2017) and later on in the southern hemisphere during the perihelion season ($L_s$ = 210° to 310°) and at different locations in the latitude band from 5°S to 40°S (Clancy et al. 2021). Clancy et al. (2009) named these features Perihelion Cloud Trails (PCT), and they reported that they have lengths ∼ 200 − 1000 km, are oriented from W to WSW, and form at high altitudes ∼ 40-50 km, based on images taken by MARCI/MRO.

However, VMC images, which cover an ample range of local times, have shown that this phenomenology occurs in many more places and in other periods of the Martian year (Figure 12). For example, in the northern hemisphere, westward-oriented elongated clouds have been observed starting at the crater Lyot (50.5°N) at $L_s$ = 340°, with lengths ∼ 1500 km (Fig. 12a). Also in the north, a series of parallel filaments oriented north-east have been observed in the Alba Patera region (45°N) at $L_s$ = 218° - 285° (Fig. 12d-e). Long filaments, some of them as long as ∼ 2600 km, have also been observed in relation to the complex morphology of the North Polar Hood (61°N and $L_s$ = 301°) (Fig. 12f). In the south, series of filaments with lengths up to ∼ 2500 km have been observed emanating from the southern wall of the Hellas basin at 41°S and at $L_s$ = 26°-40° (Fig. 12b), and at 40.5°S in the Solis Planum Highlands, north of the Argyre Planitia, with a length of ∼ 1700 km at $L_s$ = 34° (Fig. 12c).

The similarity of those elongated clouds with the AMEC, including in some of the cases the presence of a well-defined head in morning hours (Fig. 12a,c,d), suggests the action of a similar mechanism. However, taking into account the different shapes, heights and sizes of the orographic formations that originate the clouds, together with the diverse local atmospheric conditions in relation to those at Arsia Mons, a common mechanism is unlikely. Therefore, the formation mechanism of these other elongated clouds remains to be explored.

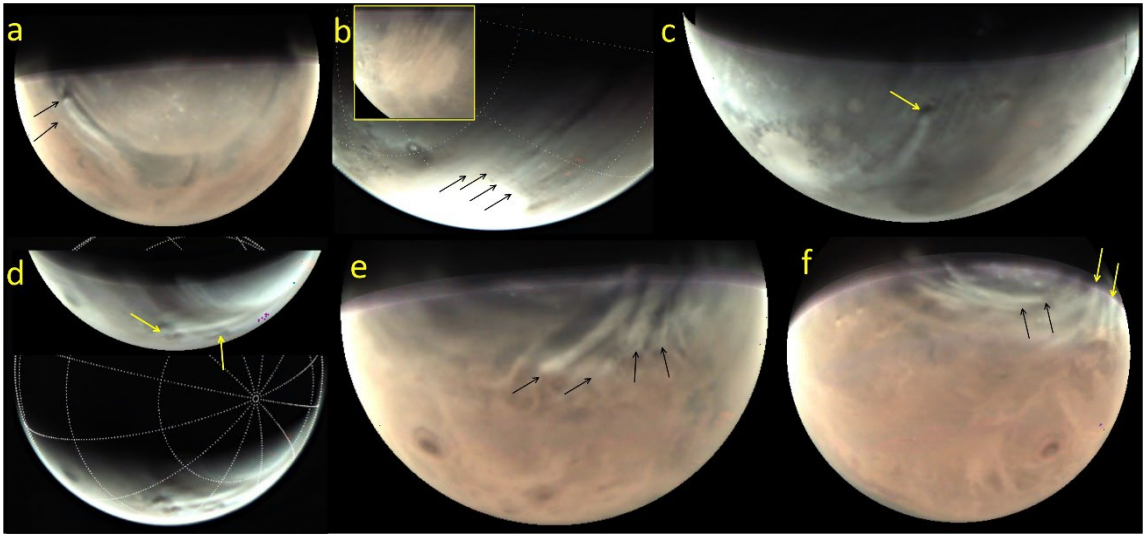

**Figure 12.** *VMC images showing elongated clouds (also called trails or filaments, marked with arrows) in the Northern hemisphere (a, d, e, f) and in the Southern hemisphere (b, c). (a) Crater Lyot on 31 December 2020 (33.7°E, 50.5°N, 7.2 hr LTST, MY 35, $L_s$ = 340°). A second parallel and narrower trail has its origin at 39.6°E, 44.4°N. (b) Series of filaments in Hellas basin (overexposed images) on 3 April 2021 (62.3°E, 41.3°S, 10.5 hr LTST, MY 36, $L_s$ = 26°). The inset show the filaments over Hellas. (c) Solis Planum Highlands on 21 April 2021 (270.3°E, 40.5°S, 7.5 hr LTST, MY 36, $L_s$ = 34°). (d) Alba Patera region on 29 April 2022 (257.9°E, 43.4°N, 9.4 hr LTST, MY 36, $L_s$ = 217°). (e) Alba Patera region on 7 August 2022 exhibiting series of parallel filaments (longest at 252.6°E, 47.5°N, 11.3 hr LTST, MY 36, $L_s$ = 280.8°). (f) Series of filaments in the North Polar Hood on 10 September 2022 (the long filament in the center 215°E, 61.1°N, 11.8 hr LTST at the head of the filament, MY 36, $L_s$ = 301.5°). The coordinates and LTSTs correspond to the head of the filaments.*

### *Valles Marineris ground fogs and dust layers*

Hazes and dust layers on the Valles Marineris have been observed by VMC, HRSC, OMEGA, and PFS in different circumstances and seasons of the Martian year (Figure 13). Inada et al. (2008) analyzed these aerosols based on OMEGA hyperspectral imaging in May 2004 at $L_s$ = 38°-45° and concluded that they were made of dust. However, Möhlmann et al. (2009) analyzed the same period with PFS and they determined that the hazes are composed of water-ice particles, probably mixed with the red dust detected by Inada et al. (2008) that acted as condensation nuclei. The water-ice fogs in the Valles Marineris at that season result from the low temperatures reached at depths of -5 km during that period, and they occupy ~ 2 km in thickness. Deposits of dust layers, revealed by their higher albedo compared to the surrounding terrains, can also cover the floor of Valles Marineris. They are observed during the dusty season, related to the presence of dust storms that inject and deposit the dust into the Valley.

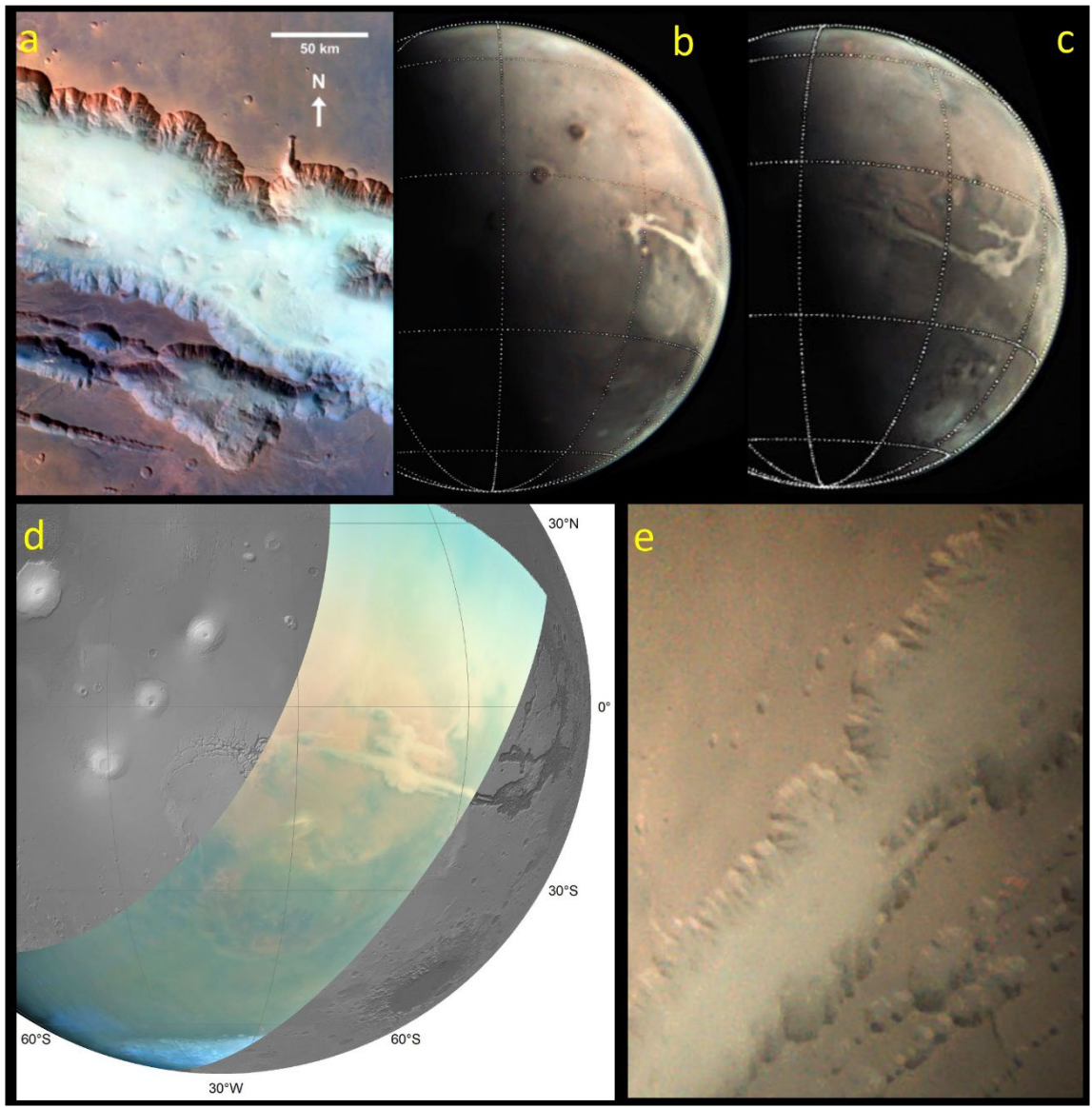

**Figure 13.** *Aerosols in Valles Marineris. (a) Water-ice fog near surface, HRSC image on 25 May 2004 (L$_s$ = 38°); (b)-(c) Water-ice fogs, VMC images on 26 July 2008 (L$_s$ = 139°) with North up and East at right; (d) Dust spreading in the Valley from the Global Dust Storm GDS 2018, HRSC image on 26 July 2018 (L$_s$ = 220°); (e) Dust near the Valley surface, VMC image on 12 November 2018 (L$_s$ = 287°).*

### *Lee waves and Gravity waves*

Gravity waves are atmospheric oscillations in which buoyancy acts as the restoring force (Fritts and Alexander 2003). As such, they can only exist in a stably-stratified atmosphere. Gravity waves, arising in a wide range of spatial and temporal scales, affect the atmospheric dynamics, transporting energy and momentum from the lower atmosphere to higher altitudes and influencing, when undergoing breaking or non-linear propagation, the mean circulation and thermal structure of these higher altitude regions (Gilli et al. 2020; Heavens et al. 2020). On Mars, the mechanisms of generation of gravity waves can be associated with surface features (orographic forcing), convective processes, and wind shear (Fritts and Alexander 2003). In addition, vertically propagating gravity waves can create localized cold pockets, promoting the condensation of $CO_2$ ice clouds, detected with the OMEGA instrument (Määttänen et al. 2010; Spiga et al. 2012). Trapped lee waves are stationary gravity waves that form in the lee side of an orographic obstacle, when atmospheric conditions trap the wave between two reflective layers, resulting in a perturbation that extends long distances with little attenuation.

If temperature minima are low enough to allow condensation, gravity waves become visible in images as trains of parallel clouds (as is the case in Earth's atmosphere). Orographic lee waves made of water-ice clouds are observed departing from crater rims of Mars mostly in high latitudes (50°-80° N and S) and during the non-summer seasons (Clancy et al. 2017). They extend long distances $\sim 100 - 500$ km from the obstacle with wavelengths in the range 3–80 km and extending vertically between 10 and 20 km (Clancy et al. 2017). Imagers OMEGA and HRSC have observed lee wave clouds mostly during fall and winter on the northern Planitias, around the North Polar Hood (Figure 14). Nadir observations of OMEGA have imaged such clouds on the rough region of Tempe Terra and extending to Acidalia Planitia, as well defined wave-packets, with brighter crests and darker droughts (Brasil et al. 2022). HRSC have observed lee wave clouds departing from Phlegra Mons, a mountain located at the north of the Elysium Province (also reported recently by Ogohara et al. (2023), based on MGS/MOC images). In addition, partly coalescing lee wave clouds, some extending over larger areas, have been observed by HRSC emerging from numerous impact craters at Noachis Terra, Terra Cimmeria, and Solis Planum. When close to pericenter, VMC can also image gravity waves. They have been observed in Lyot and nearby craters, and in craters in northeast arcadia (Figure 14b), forming trains of waves extending $\sim 200$ km with wavelenths of 30-40 km. These trains sometimes interfere, giving rise to complex patterns. Identifying horizontal disturbances by gravity waves in the absence of cloud formation is challenging; however, the OMEGA spectrometer enabled such identification: Spiga et al. (2007) mapped pressure disturbances caused by gravity waves and Altieri et al. (2012) obtained horizontal maps of oxygen dayglow in which gravity wave trains could clearly be identified and confirmed by mesoscale modeling.

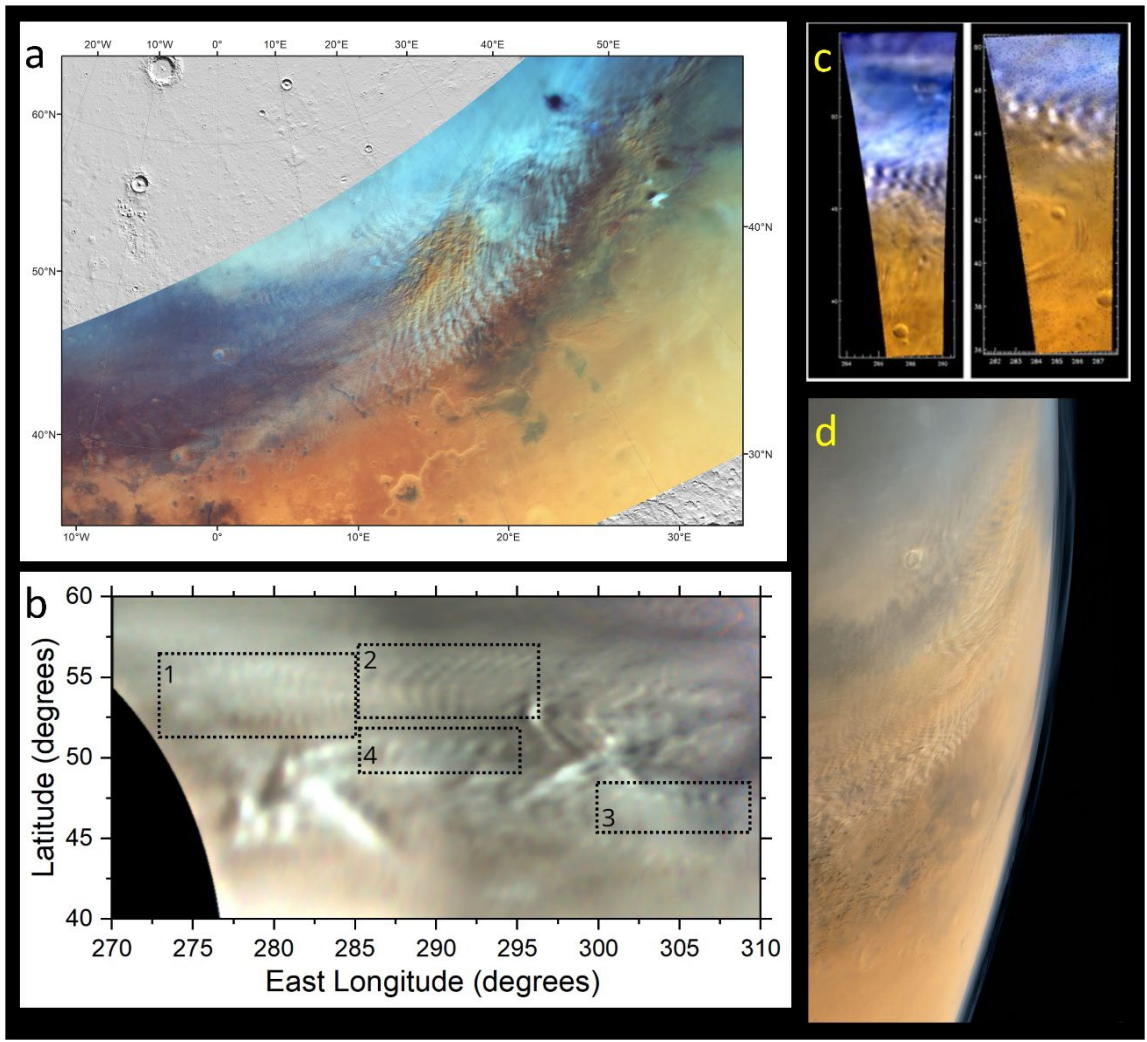

**Figure 14**. *Atmospheric gravity waves detected with the HRSC, VMC and OMEGA cameras. (a) HRSC images of a gravity wave field traced by water-ice clouds mixed with dust on 15 December 2020; (b) Gravity waves and elongated clouds in a composite projection of three VMC images taken 9 December 2022. Wave trains 1-2 and 4 are originated by craters, and are imaged at LTST around 15h, whereas 3 and the elongated clouds nearby are related to the rough orography of the Martian dichotomy. (c) Gravity waves observed by OMEGA in June 2007 (courtesy of B. Gondet); (d) Trains of gravity waves observed by HRSC interacting and interfering on 13 June 2013 (orbit 12,051). Image Credit: ESA/DLR/FU Berlin/J. Cowart, CC BY-SA 3.0 IGO*

## 6b. Convection

Convection develops on Mars as a result of the combined influence of mesoscale circulations and planetary boundary layer dynamics. Possible convective motions on Mars are dynamically similar to their terrestrial counterparts, although the forcing mechanisms of convection on Mars is radiatively-controlled rather than based on latent heating as is the case on Earth. We distinguish between what is considered as ``shallow'' convection associated with convective plumes developing in the planetary boundary layer on Mars and "deep" convection associated with convective plumes rising at higher altitudes from the surface.

Mars Express has allowed to study idiosyncratic cases of possible convective motions. Määttänen et al. (2009) used a combination of OMEGA and PFS observations to retrieve the properties of dust particles and to assess the dust optical thickness of the storm confined in the first one-two scale heights of the atmosphere. Spiga et al. (2013) performed mesoscale simulations which suggested that such an observed storm in the afternoon would induce strong, local, radiative warming of the Martian atmosphere causing convective updrafts reaching 30 to 40 km altitudes that the authors named ``rocket dust storms''.

Mars Express VMC, combined with HRSC images, allowed to study dust convection in textured dust storms (Sánchez-Lavega et al., 2022a). These were representative local-to-regional dust storms showing cumulus-like morphologies -- and/or flushing -- large fronts of lofted dust particles resulting from baroclinic instability (Figure 9). The cellular patterns had typical sizes and heights (in the range 5-10 km) that overlap with planetary boundary layer features.

## 6c. Dust devils

Dust devils (DD) are vortices of convective origin, part of daytime Planetary Boundary Layer turbulence, that lift and inject columns of dust into the atmosphere, sometimes up to heights of several kilometres (Balme and Greeley 2006). They can be observed from orbit in high-resolution images, directly when the size of the vortex and the dust plume projected on the surface is sufficiently large (up to hundreds of metres), and also by the trace they leave on the surface when the dust is stirred up (Kahre et al. 2017 and references therein, Perrin et al. 2020). Dust devils have been observed at all latitudes (except in the permanent frozen poles, poleward of $\sim 80°$), longitudes, seasons, and even altitudes, from the tops of the volcanoes to the depths of Hellas basin. Mars Express HRSC camera has captured a large number of active DDs (Figure 15), about 50% in each hemisphere (Stanzel et al. 2008). From measurements of about 200 DDs, these authors find dust-column altitude is in the range $\sim 0.3$-4.4 km, and an average diameter of dust clouds $\sim 230$ m, with most diameters in the range $\sim 50$-650 m and the largest 1.6 km (one of the largest dust devils ever imaged from orbit on Mars). DDs were observed between 11 and 16:30 hr LTST, with typical lifetimes of $\sim 5$-30 minutes. Stereo camera images allowed the measurement of translation velocities, in the range $\sim 1$ to 59 ms$^{-1}$ (average 24 ms$^{-1}$), values that probably correspond to wind speeds above the surface (Stanzel et al. 2006, 2008; Reiss et al. 2011). Spectacular images of DD tracks have also been captured by CASSIS (Colour and Stereo Surface Imaging System) onboard the Exomars Trace Gas Orbiter. The images show the streaks that the DDs leave behind their course after removing the surface material (*).

(*) https://www.esa.int/ESA_Multimedia/Images/2019/03/Dust_devil_frenzy

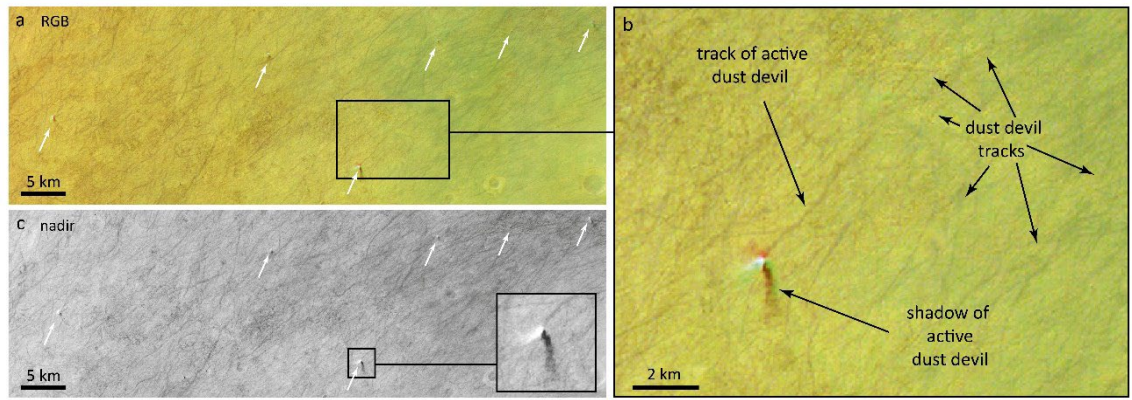

**Figure 15.** *Dust devils and dust devil tracks in Aonia Terra, southwest of Argyre Planitia observed by HRSC on September 23, 2020. (a) RGB color composite of orbit h21145_0000 showing six dust devils (white arrows) traveling NE-SW. The thin black lines are dust devil tracks. (b) Close-up image of the area outlined by the black box in panel a, showing one larger dust devil, its shadow and track as well as older dust devil tracks. Due to the time shift between the acquisition of the three color channels and due to the fact that the dust devil is moving during that time, the devil and its shadow are displaced in red, green and (faint) blue and only in the overlapping area the dust devil appears white and its shadow black. Such color offsets in composite images provide evidence of the motion of dust devils observed by HRSC. (c) Nadir image of the same orbit. This black and white view is shown as a reference without color offset. Enlarged dust devil (black box and inset) is the same as the one in panel b. Its diameter is about 134 m and the shadow length is about 1.1 km. The shadow is used to calculate the dust devil's height. The unusual yellow-greenish tint of the Martian surface in this observation is due to the atmospheric conditions at the time the image was taken.*

## 7. Conclusions and Perspectives

Mars Express VMC, HRSC and OMEGA cameras have made it possible to observe the clouds and dust that trace some of the phenomena described above from a new perspective, which equally inspired a large body of work both in data analysis and numerical modelling. In particular, the polar orbit of MEX allows precise observations of the vertical structure of the aerosol layers observed at the limb (Sánchez-Lavega et al. 2018a), which can reach heights ∼ 90 km or higher above the surface. The cameras make it possible to monitor large regions of the limb at different local times. They also allow for the study of the terminator of the planet in twilight conditions, where the structure and rapid evolution of the highest aerosols can be analyzed (Hernández-Bernal et al. 2021b). In the case of twilight observations, the highest resolution images also allow measuring the displacements of individual cloud elements to obtain the wind velocity, difficult to evaluate by other procedures at that time of the day. Details of all these studies can be found in the accompanying paper by Määttänen et al. (this issue).

We have shown that VMC, a simple and cheap camera on-board MEX, although initially developed only for technological purposes, provides images of high scientific value, in addition to its well proven educational value, outreach and citizen science (see Wilson et al. this issue). This scientific potential is greatly enhanced when combined with images taken by other cameras on MEX that provide high spatial resolution (HRSC) and broad spectral sensitivity (OMEGA), and with cameras from other orbiters, particularly those imaging Mars continuously, as MARCI/MRO from its sun-synchronized orbit, which allow improving the spatial resolution and time coverage of different phenomena (Ordóñez-Etxeberría et al. 2022).

Many of the meteorological phenomena presented here are targets for future studies. Future research objectives include the analysis of the atmospheric disturbances tracked in the images and their possible imprint on the meteorological measurements at Jezero crater made with the MEDA instrument on board the rover Perseverance (Battalio et al., 2022; Sánchez-Lavega et al. 2023). We also want to complete the study of the clouds at twilight, extending their analysis in time in order to complete the statistics of their spatial and temporal occurrence while measuring the motions of individual elements to retrieve the velocity distribution in morning and evening hours, the areographic location (latitude and longitude) and seasonal epoch (Hernández-Bernal et al. 2022b). Another objective is the study of the elongated clouds (filaments) generated by different types of orographic structures and their comparison with AMEC in order to determine if the mechanism underlying their formation is similar or not. Finally, in connection with this last phenomenon and with the study of clouds at twilight, we intend to explore the zonal alignments of clouds and hazes observed around Martian sunrise and sunset. VMC images these phenomena at high temporal resolution, but in low spatial resolution. HRSC double brooms have the potential to combine high spatial and temporal resolutions, and it might be able to provide new observational insights into the dynamics of these clouds. Time lapses made from VMC observations and published in the supporting information of Hernández-Bernal et al. (2021a) are an example of the rich dynamics to be explored, which are probably governed in part by non-linear gravity wave processes.

We remark that the best global and simultaneous measurements of the full-disk are provided by missions with high altitude orbits (MEX, MAVEN, MOM, EMM, Tianwen-1). Also, continuous tracking of clouds and storms is key to the understanding of the global dynamics, therefore the capacity to perform multi-temporal images is absolutely necessary to monitor winds, making the wide angle cameras (VMC, MCC) very important despite their low spatial resolution. Unfortunately, some of the cameras (MOM/TIS, Singh 2015; Tianwen-1 MORIC, Yu 2020 and HIRIC, Meng 2021) are still largely unexploited and have not yet produced significant science results for atmospheric phenomena (no refereed publications yet available). We can conclude that high altitude, simultaneous, global and continuous coverage of the planet, as proposed by many future mission studies and proposals (Montabone et al. 2022) would be key for the future improvement of our understanding of Mars meteorology and dynamics.

**Acknowledgements**

ASL and TdR have been supported by Grant PID2019-109467GB-I00 funded by MCIN/AEI/10.13039/501100011033/ and by Grupos Gobierno Vasco IT1742-22. AS acknowledge support from CNES. JHB and EL were supported by ESA Contract No. 4000118461/16/ES/ JD, Scientific Support for Mars Express. PM was supported by ESA Mars Wind and Wave Mapping contract No. 4000138190/22/ES/CM. We acknowledge support from ESA through the Faculty of the European Space Astronomy Centre (ESAC) - Funding reference ESAC-531. The authors thank the International Space Science Institute (ISSI) in Bern, Switzerland for the support. We thank Michael Battalio and an anonymous reviewer for their helpful comments that improved the manuscript.

**Declarations**
**Competing Interests** The authors declare no competing interests.